\documentclass[preprint,nocite]{epl}
\newcommand{\bb}{\begin{equation}}
\newcommand{\en}{\end{equation}}

\title{Spreading of Latex Particles on a Substrate}
\author{A.W.C. Lau\inst{1,2}, M. Portigliatti\inst{1}, E.
Rapha\"{e}l\inst{1}, and L. L\'{e}ger\inst{1}}
\institute{
   \inst{1} Laboratoire de Physique de la Mati\`{e}re Condens\'{e}e (UMR
   CNRS 7125, FR CNRS 2438),
Coll\`{e}ge de France, 11 Place Marcelin Berthelot,
75231 Paris Cedex 05, France\\
   \inst{2} Department of Physics and Astronomy, University of
Pennsylvania, Philadelphia, PA 19104
}

\begin{document}

\maketitle

\begin{abstract}
We have investigated both experimentally and theoretically the
spreading behavior of latex particles deposited on solid substrates.
These particles, which are composed of cross-linked polymer chains,
have an intrinsic elastic modulus. We show that the elasticity must be
considered to account for the observed contact angle between the particle
and the solid substrate, as measured through atomic force microscopy
techniques. In particular, the work of adhesion computed within our model can be
significantly larger than that from the classical Dupr\'{e} formula.
\end{abstract}

\section{Introduction}

Adhesion phenomena are of great importance in science and
technology\cite{nature,kendall}.
The classical approach to describe the adhesion between two elastic bodies
under a compressive force is given by the Johnson, Kendall, and Roberts
(JKR) theory\cite{jkr}. It is essentially an extension of the Hertz theory
of elastic contact, allowing for a tensile stress to develop inside contact
area to account for the effect of adhesion.
For an adhesive sphere of radius $R$ in contact
with a flat solid substrate, the JKR theory predicts
that the deformation of the sphere at the center of the contact area
is $\delta= {a^2 \over R } \left [ 1 - {2 \over 3} \left ( {a_0 \over a
}\right )^{3/2} \right ]$, where $a$ is the contact radius
given by
$a^3 = \frac{3R}{4K} \left[F + 3 \pi R W + \sqrt{6 \pi R W F +
(3 \pi R W )^2}  \right]$. Here $a_0 = (9 \pi R^2 W /2K )^{1/3}$ is
the contact radius at
zero force ($F=0$),  $W$ is the thermodynamic work of adhesion\cite{li1},
$K$ is the rigidity of the sphere $K \equiv E_B/(1-\nu^2)$, $E_B$ is the Young's modulus, and
$\nu$ is the Poisson's ratio. If $W =0$, the JKR theory
reduces to the Hertz result for elastic contact of a nonadhesive
spherical particle with a planar substrate, for which $\delta = a^2/R$\cite{landau}.

Recent attention has focussed on the adhesion properties of rigid spheres placed on
a soft matrices where large deformation occurs.  Rimai {\em et al.} \cite{rimai}
have studied this problem experimentally while Maugis \cite{maugis1} has extended the JKR analysis
to large contact radii.  In this paper, we consider the ``inverse'' problem of soft spherical particles
adsorbed on a rigid planar substrate, on which they tend to spread
and deform under the effect of spreading forces.  Our first aim is to present
a series of experiments conducted by atomic force microscopy
techniques to characterize the shape of isolated latex particles deposited
on a clean silicon wafer surface.  Using AFM, we determine the equilibrium height of the deformed
latex particles, and to deduce a contact angle.  These experiments demonstrate
that the final equilibrium shape of the particle does not depend only on the surface tensions,
like for a liquid drop, but is greatly affected by the elastic modulus of
the particle.  Since in these experiments the particles deformation is
rather important, one cannot use the JKR theory which assumes that $a \ll R$. Therefore,
in a second step,  we present a simple model which predicts the contact angle
of an elastic sphere deformed under the effect of spreading forces even for
rather large deformation (but still assuming linear elasticity).  Very recently,
Joanny {\em et al.} \cite{joanny} have considered a similar problem of spreading
of a ``cylindrical'' droplet of gel in a regime where the role of surface tension
is not important.  Here, we consider spherical geometry and explicitly take the surface
tension into account, which are more appropriate for our experiments.
We point out that a more accurate estimate of $W$ should take into account
the elastic energy, which is not contained in the Dupr\'{e} formula,
$W = \gamma_{s} ( 1 + \cos \theta )$, which relates the surface tension of
a liquid drop $\gamma_s$ and the contact angle $\theta$ to $W$.

\section{Experimental Investigation of the Spreading Behavior of
Elastic Spheres}

\begin{table}
{\scriptsize
\begin{center}
\begin{tabular}{ccccccc}
\hline
Latex & $T_g$ & GF ($\%$) & $N$ ($\mu$mole{/g})
& $R$ ($\mbox{nm} \pm 5\,\mbox{nm}$) & $E_B$ ($\mbox{Mpa}$) & $\gamma_s$ (mJ/m$^{2}$) \\
\hline
SB-(-2)-75 & -2 & 75 & 210 & 175 & 0.63 & 53 \\
SB-(11)-75 & 11 & 75 & 235 & 167 & 1.35 & 46 \\
SB-(28)-75 & 28 & 75 & 225 & 167 & 1.92 & 48 \\
SB-(-2)-92 & -2 & 92 & 310 & 148 & 0.6  & 54 \\
SB-(-8)-43 & -8 & 43 & 334 & 164 & 2.02 & 48 \\
SB-(-7)-68 C- & -7 & 68 & 190 & 138 & -- &   45\\
SB-2-68 C+ & 2 & 68 & 427 & 148 & -- &  56\\
\hline
\end{tabular}
\end{center}
}
\caption{Characteristics of the latex particles: the glass transition temperature $T_g$,
gel fraction (GF), number of acidic function per unit area $N$, the diameter $R$,
and bulk elastic modulus $E_B$ and surface tension $\gamma_s$.}
\label{data1}
\end{table}

\begin{table}
{\scriptsize
\begin{center}
\begin{tabular}{cccccc}
\hline
Latex & $R$ ($\mbox{nm} \pm 5\,\mbox{nm}$) &  $h$ (nm) &
$2 a$ (nm) & $\theta$ ($^\circ$) & $W_D$ (mJ/m$^2$) \\
\hline
SB-(-2)-75 & 175 & 26 & 640 & 11 & 106 \\
SB-(11)-75 & 167 & 39 & 610 & 22 & 88  \\
SB-(28)-75 & 167 & 49.5 & 565 & 32 & 89 \\
SB-(-2)-92 & 148 & 46  & 398 & 34 & 99 \\
SB-(-8)-43 & 164 & 15.8 & 695 & 6 & 95 \\
SB-(-7)-68 C- & 138 & 37 & 365 & 27 & 85 \\
SB-2-68 C+ & 148 & 23 & 460 & 12 & 111 \\
\hline
\end{tabular}
\end{center}
}
\caption{Geometrical characteristics of the latex particles adsorbed on silica, as determined through AFM,
including the height of the adsorbed particles $h$, the contact radius $a$, the contact angle $\theta$
and Dupr\'{e} energy $W_D$.}
\label{data2}
\end{table}

The latex particles used in the present study are formed by a soft
core of partially crosslinked styrene-butadiene copolymer
molecules, surrounded by a thin stiffer shell made of carboxylic
co-monomers. The ratio of styrene over butadiene co-monomers in
the core allows one to adjust the glass transition temperature $T_g$
of the latex, while the elastic modulus can be varied by adjusting
the degree of crosslinking of the core attained during the
formation of the particles. In fact, it is not easy to
quantitatively determine the elastic modulus of an isolated latex
particle. Two quantities can be used to qualitatively estimate
this elastic modulus. First, the gel fraction, GF, represents the
ratio of insoluble species remaining after swelling the particles
in a good solvent of the styrene butadiene copolymers. The larger
GF, the higher is the degree of crosslinking and the higher the
elastic modulus. However, the correspondence between them is not
quantitative, because the crosslinking reaction is accompanied by
chain scissions. Second, the bulk elastic modulus $E_B$ of
macroscopic films which can be formed from slowly evaporating the
water of a latex suspension can easily be measured. These films
are made of closed packed particles, adhering together by their
shells. It is, however, not totally obvious that the bulk elastic
modulus thus measured corresponds exactly to the elastic modulus
of an isolated particle. The characteristics ($T_g$, GF, $E_B$, radius)
of the latex used are reported in Table \ref{data1}.

\begin{figure}
\onefigure[scale=0.4]{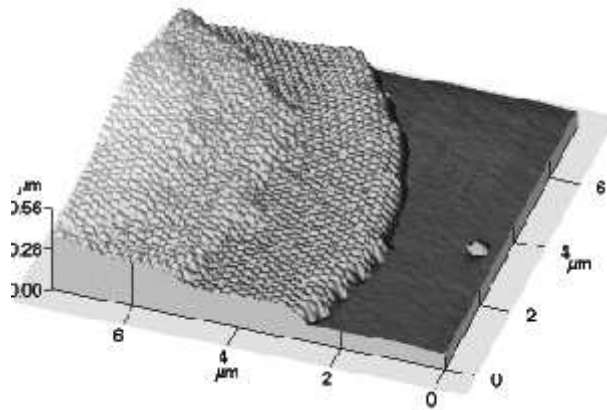} \caption{AFM image of the edge of
a drop of latex (SB-(-2)-75) deposited on a silica surface and
dried: an isolated particle is visible on the right of the image.}
\label{fig1}
\end{figure}

\begin{figure}
\onefigure[scale=0.6]{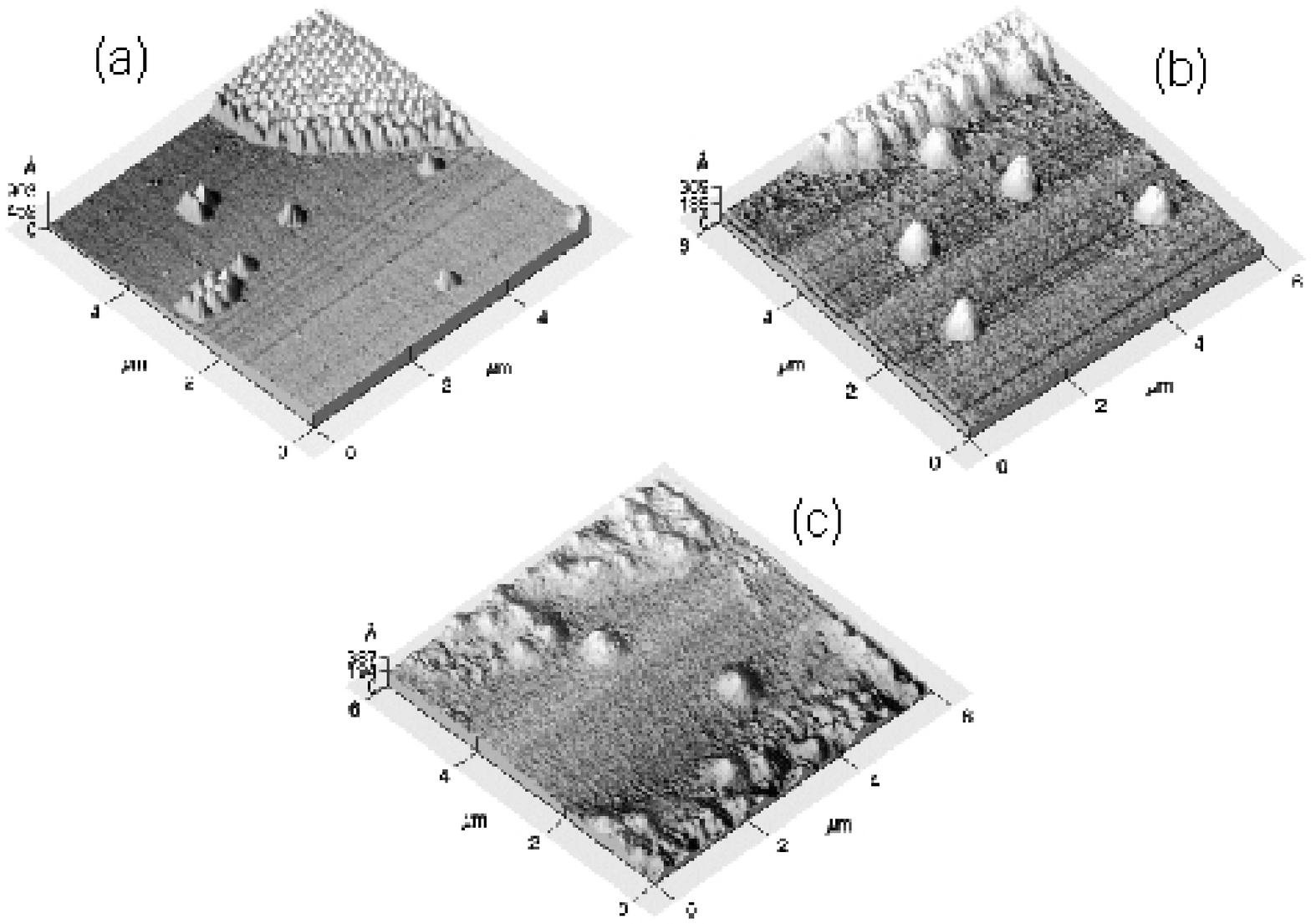} \caption{Latex particles adsorbed
on a silica surface (AFM image, contact mode under a constant load
of $5\,\mbox{nN}$.) (a) Latex SB-(-2)-92, (b) Latex SB-(-2)-75,
(c) Latex SB-(-8)-43.} \label{fig2}
\onefigure[scale=0.5]{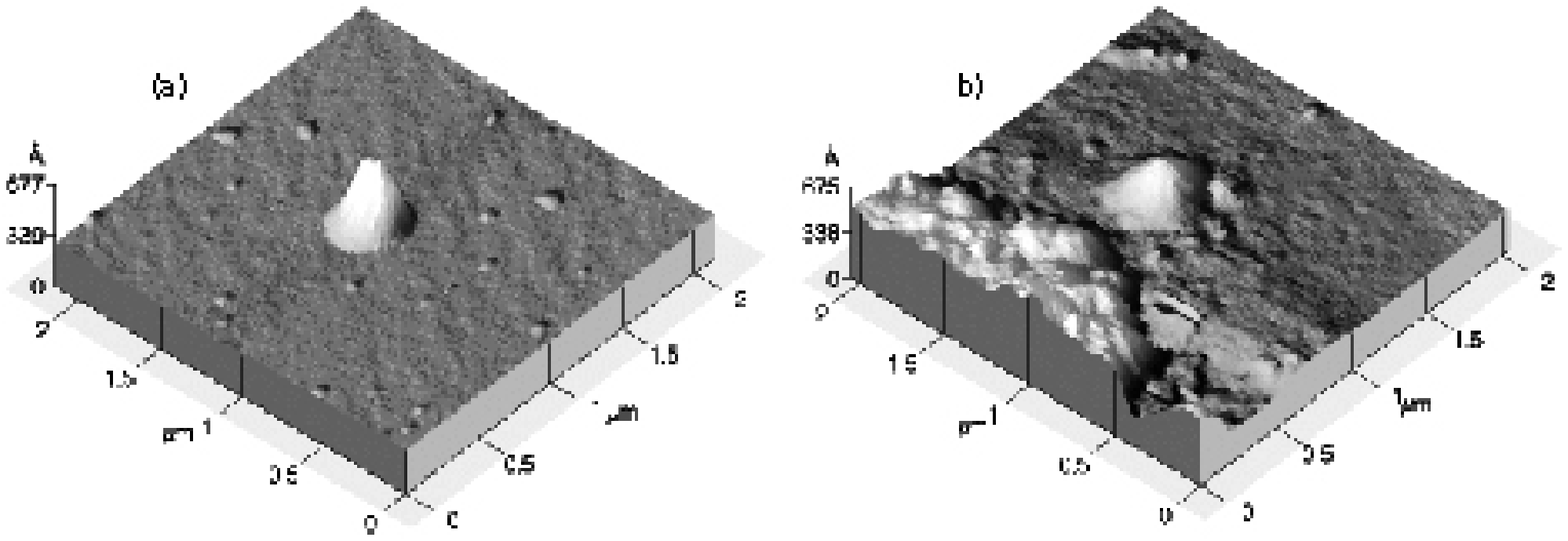} \caption{Latex particles adsorbed
on a silica surface (AFM image, contact mode under a constant load
of $5\,\mbox{nN}$.) (a) Latex SB-(-7)-68 C-, (b) Latex SB-(2)-68
C+.} \label{fig3}
\end{figure}

In order to investigate the spreading behavior of these latex on controlled
surfaces, the latex needs first to be thoroughly washed in order to extract the
surfactant molecules used to stabilize the suspension during the
synthesis. Then a drop of a highly diluted suspension is deposited
on a clean silicon wafer surface \cite{li7}. Under slow evaporation of
the water, a drop of well organized latex particles forms, with
particles arranged successively from the edge of the drop in a
closed packed monolayer, and then bi-layer, and then multilayers,
as shown in the AFM image (contact mode) of Fig. \ref{fig1}. Ahead of
this continuous drop, a few isolated particles remain (Fig.
\ref{fig1}), and can be used to characterize the spreading behavior of isolated
particles. It is important to note that while we cannot verify that the sol fraction
of the particles has been significantly extracted during the washing procedure, using
AFM we do not observe any leakage or spreading of this sol fraction ahead of the particle,
except perhaps in the case of the latex with the highest sol fraction where some leakage
become visible.  We thus think that the sol fraction only affects the viscoelastic
properties of the particles, and not significantly their spreading behavior.
In Fig. \ref{fig2}, AFM images of latex particles with
similar glass transition temperatures, similar radius and different GF are
reported. It appears clear that the softer particles are more
deformed and spread than the harder ones, while all having the
same shell should have the same spreading parameter. In Fig.
\ref{fig3}, AFM images of latex particles having the same GF and different
shells are presented, showing that, with the same elastic modulus,
the more polar shell leads to a higher deformation of the
particles. These experiments clearly demonstrate that both the
surface energy and the elastic modulus of the particles govern
their spreading behavior. It is quite easy, from AFM, to
quantitatively measure the height $h$ of the deformed particles.
Their diameter at the surface can also be measured, but this
measurement is far less accurate than that of the height, as it
needs to de-convolute from the tip radius $R_{tip}$ \cite{li8}.

From the geometrical characteristics of the deformed particles a contact
angle can be deduced: assuming that the particles take the shape
of a spherical cap, with a radius $R_{cap}$ given by
$R_{cap}= ( 4 R_0^3 + h^3 )/ (3h^2) - R_{tip}$, ($R_0$ is the
initial radius of the particle), the contact angle is given by
$\theta = \sin^{-1} \left [ { 1\over 2 R_{cap} } \, \sqrt{{2 \over 3}
\left (  {16 R_0^3 \over h} - 2 h^2\right )}\right ].$
An estimate of the thermodynamic work of adhesion can then be
deduced from the Dupr\'{e} formula.  The corresponding data are reported in Table \ref{data2}.
While the order of magnitude of $W$ appears reasonable,
and comparable to similar data extracted by Unertl {\em et
al.} on different latex \cite{li9,li10}, it seems clear, however, that
owing to the obvious influence of the elastic modulus on the degree of
spreading of the different latex we have investigated, it is not
correct to analyze this spreading without taking into account the
balance between adhesive and elastic energy. This is the aim of
the model that we develop now.

\section{The Model}

Consider an elastic sphere with radius $R$ spread on a high-energy
surface. For simplicity, we assume that the sphere deforms into a
spherical cap of radius $R_1= ( 4 R^3 + h^3 )/ (3h^2) \geq R,$
under the constraint that the volume is conserved, where $h$ is
the height of the deformed particle.  The total energy of the
system is the sum of two terms, namely, the surface energy $U_{s}$
and the elastic energy $U_{el}$.  The surface energy may be
written as \cite{nature,li1} \bb
U_{s} = - {4 \pi \over 3}\,R^2\,S\,\left ({2 R \over
h} \right ) + {4 \pi \over 3}\,R^2 \,( 3\gamma_s  + S )\, \left
({h \over 2 R}\right )^2,
\en
where $S \equiv \gamma_{SO} - ( \gamma_{SL} + \gamma_s )$ is the spreading parameter.
It is easy to check that minimization of $U_{s}$  with respect to $h$ leads
to Young's law $\gamma_{SO}  = \gamma_{SL} + \gamma_s \cos
\theta$. It describes the contact angle $\theta$ of a liquid drop
on a solid surface.
\begin{figure}
\twofigures[scale=0.35]{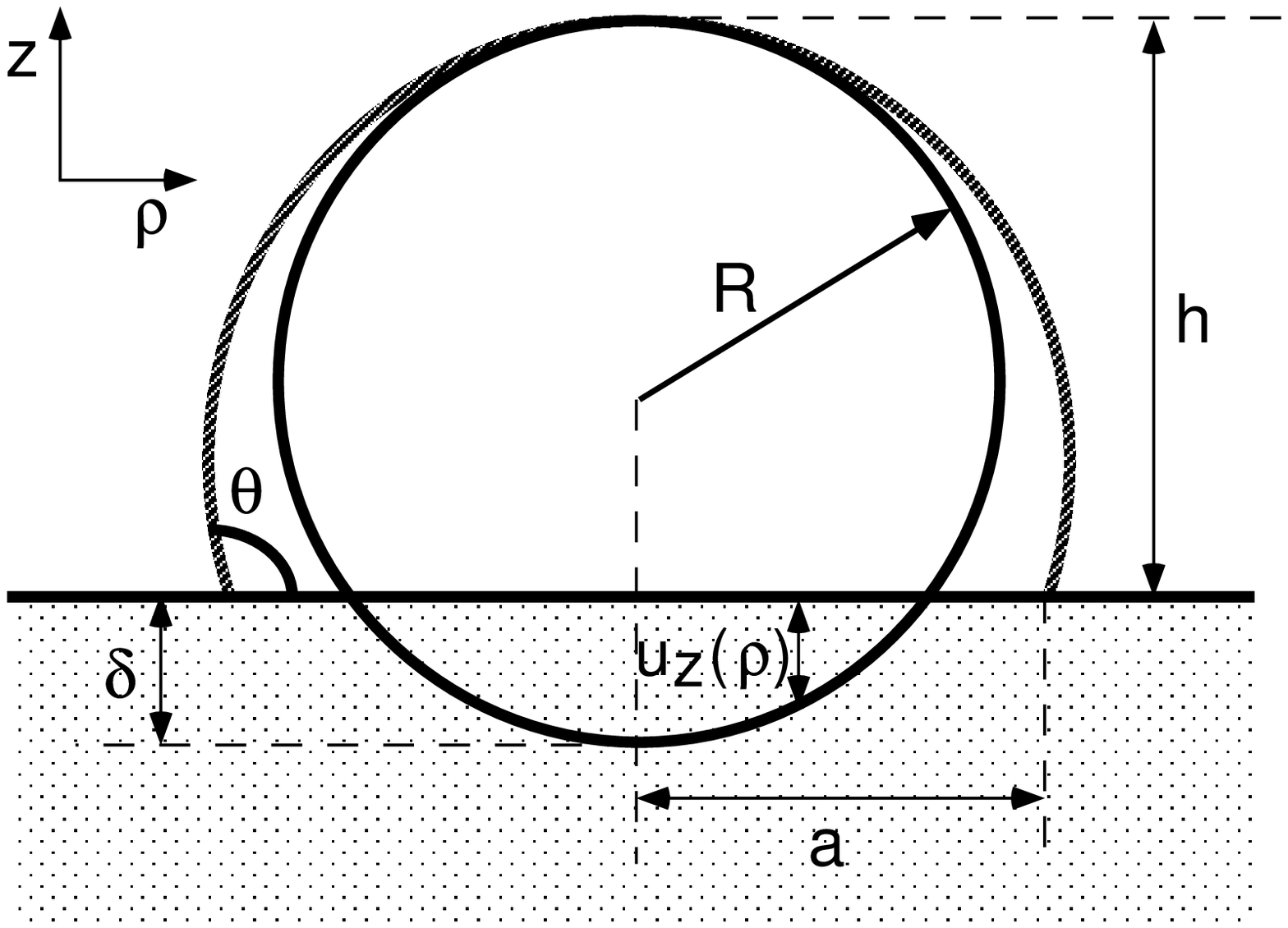}{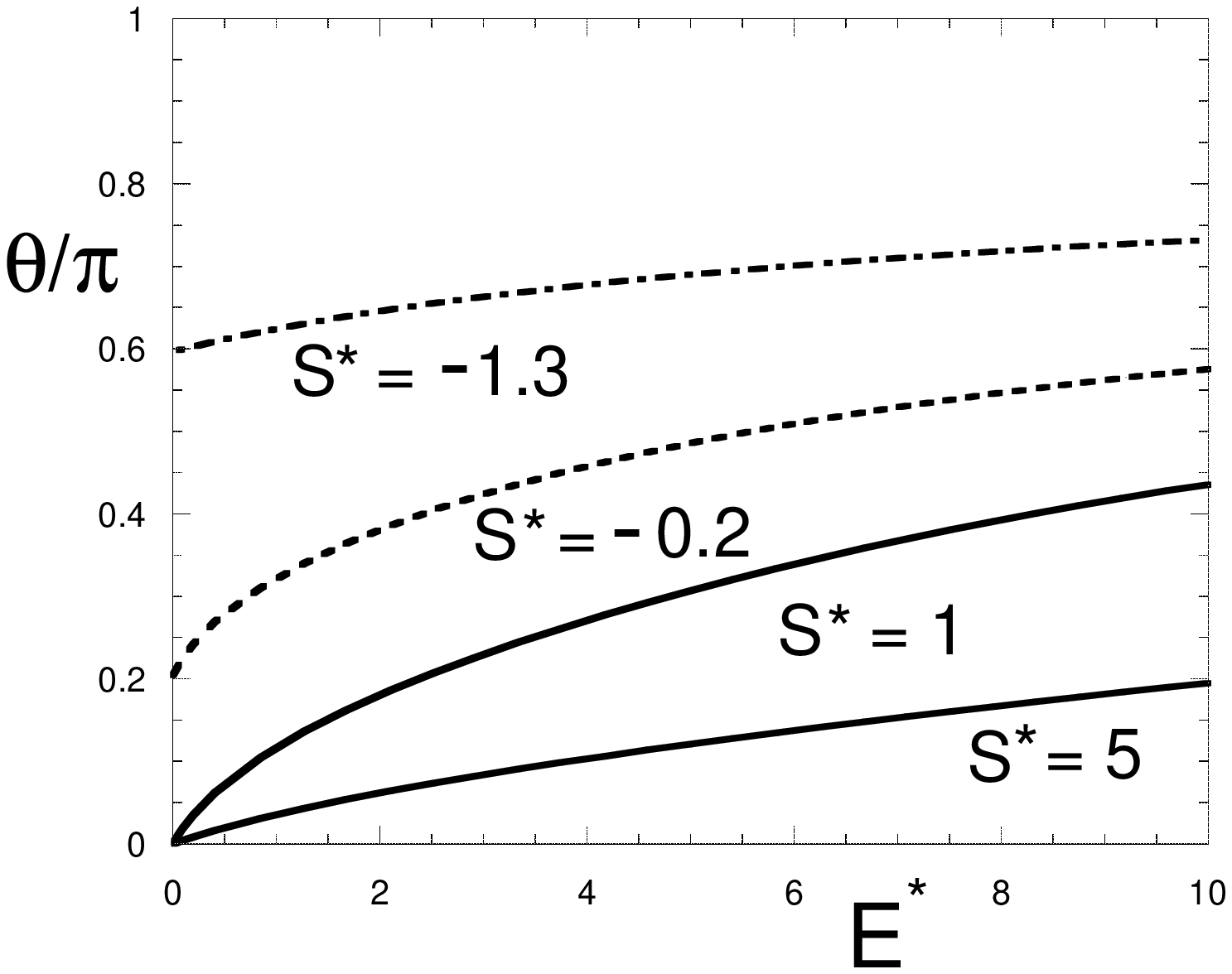} \caption{The
geometry of a deformed elastic sphere.} \label{sphere}
\caption{The contact angle as a function of the normalized elastic
modulus $E^* \equiv 2\sqrt{3} K R/(\pi \gamma_s)$ for different
values of the reduced spreading parameter $S^* = S /\gamma_s$.}
\label{contacta}
\end{figure}

For the case of the latex particles, we must take the
elastic energy stored into the deformation into account.  Unfortunately,
even with the simplifying assumption of a spherical cap deformation,
a rigorous computation for
the displacement field for such a deformation remains a difficult task.
Therefore, we must resort to a scaling picture to
obtain the elastic energy stored in such a deformation. Consider
a deformed elastic sphere shown in Fig. \ref{sphere}; the $z$-component
of the displacement field within the contact area of radius $a$ directly
follows from the geometry\bb
u_z(\rho) = \delta - R \left [1 - \sqrt{ 1 - (\rho/R)^2}\right ]
\approx \delta \left (
1 -{ \rho^2 \over 2\delta R} \right ),
\label{dis}
\en
where $\delta \equiv  2 R - h$ is the central displacement.
Note that due to volume conservation, the contact radius $a$
is determined in terms of the height of the spherical cap $h$ by:
$a^2 = {4 \over 3 } R^2 \left [ {2R \over h} - {h^2 \over
(2 R)^2} \right ].$
On the other hand, if an external stress of the form\bb
\sigma(\rho) = \sigma_0\,\left ( 1 - { \rho^2 /a^2} \right )^{-1/2}
+ \sigma_1 \, \left ( 1 - { \rho^2 / a^2} \right )^{1/2}
\,\,\,\,\hphantom{0,}\rho < a,
\en
is exerted on the surface of a semi-infinite half-space elastic
medium, it will be displaced by an amount\cite{maugis}\bb
u_z(\rho) =  {\pi a \over K} \left [ \sigma_0 + {1 \over
2}\,\sigma_1 \left ( 1 - { \rho^2 \over 2\,a^2} \right ) \right ],
\label{dis2}
\en
where $K \equiv E_B/( 1-\nu^2)$ is the rigidity.
Comparing Eq. (\ref{dis2}) to Eq. (\ref{dis}), we obtain
$\sigma_0 = {K \over \pi}\left ( {\delta  \over a} - {a \over R}
\right )$ and $\sigma_1 = {K \over \pi}\,\left ( {2a \over R} \right )$.
Therefore, the elastic energy follows from\bb
U_{el} = {1 \over 2} \int d^2{\bf x} \,  u_z({\bf x}) \sigma_z({\bf x})
= K \left [ \delta^2 a - {2 \over 3 } {\delta a^3 \over R}
+ {1 \over 5} {a^5 \over R^2} \right ] = {8 \over \sqrt{3}}\,K R^3
\,\Phi(h/2R),
\label{energy}
\en
where in the last line, we have made used of the volume conservation
constraint and $ \delta = 2 R - h$.  Note that the elastic
energy is a function of the height $h$ only with a scaling function given
$\Phi(x) \equiv \sqrt{x^{-1} - x^2} \left [ (1 -x)^2 - {4 \over 9} ( 1
-x) \left (  x^{-1} - x^2 \right ) + {4 \over 45 }\left (  x^{-1} - x^2
\right ) \right ]$, which has the following asymtotics: $\Phi(x) \sim
x^{- 5/2}, \,x \ll 1$ and $\Phi(x) \sim ( 1 - x )^{5/2},\, x \sim 1$.
We have made the following assumptions in deriving Eq. (\ref{energy}): First,
we have employed linear elasticity theory as in the Hertz and JKR
theory. Secondly, in calculating the displacement field in Eq. (\ref{dis2}),
we have made use the results from the half-space elastic medium.  For
small deformation, these assumptions are certainly justified, and the scaling
of free energy Eq.(\ref{energy}) with $h$ is in fact identical to the Hertz theory \cite{maugis}.
For large deformation, it can be argued that Eq. (\ref{energy}) should at least give the right
scaling with $h$. To see this, consider an elastic ball which develops a contact radius
of $a^2 \sim R^3/h$ for large deformation; the strain is then of
the order of $\epsilon \sim a/R$. In a linear theory, the elastic energy must scale like
$U_{el} \sim a^3 \epsilon^2 \sim h^{-5/2}$, as obtained above.
We note that in contrast to Hertz theory, Eq. (\ref{energy})
diverges as $h \rightarrow 0$. As we shall see below, even with this
crude estimate of the elastic energy, our results compare quite well with the experimental data.

\section{Results and Discussion}

\begin{figure}
\twofigures[scale=0.35]{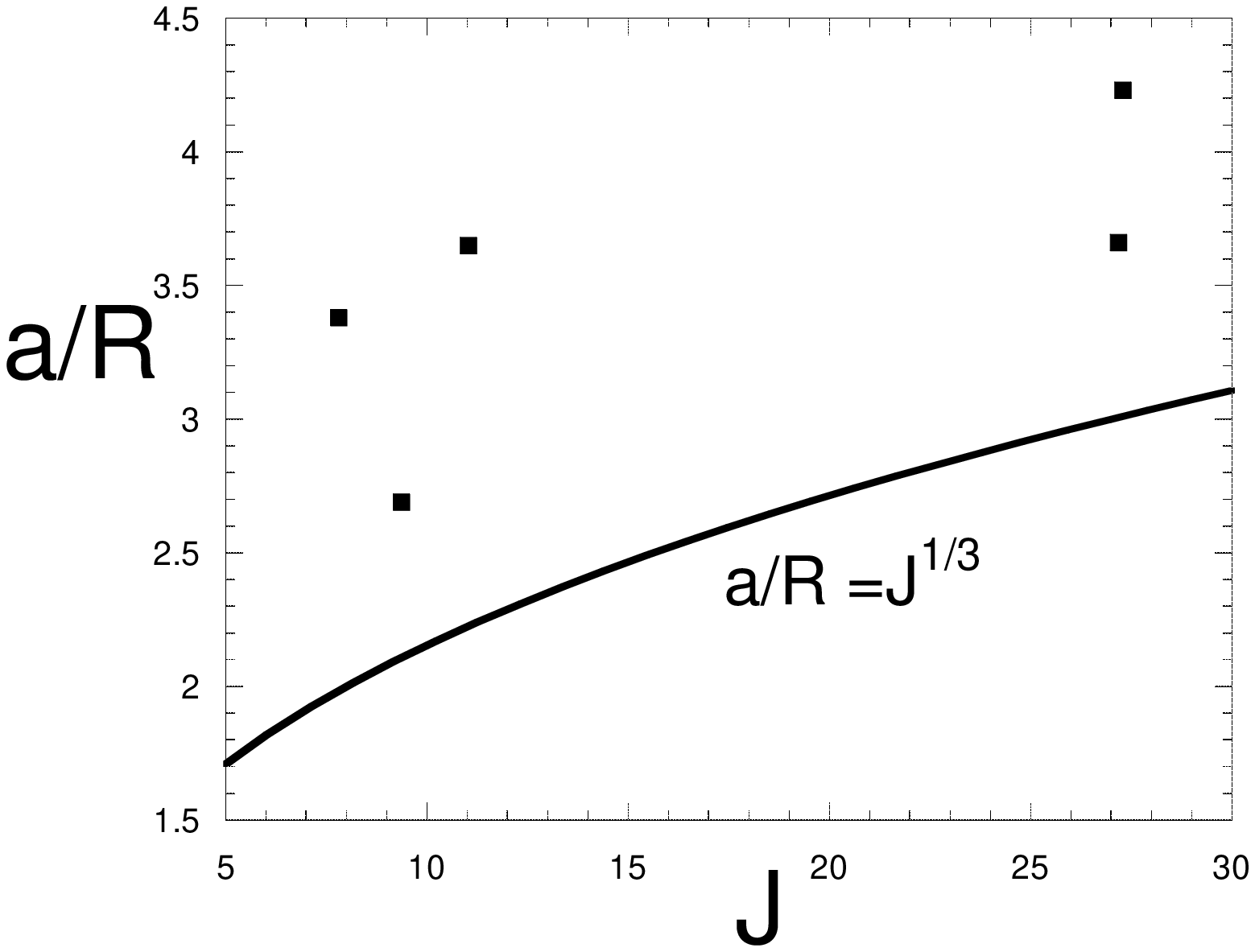}{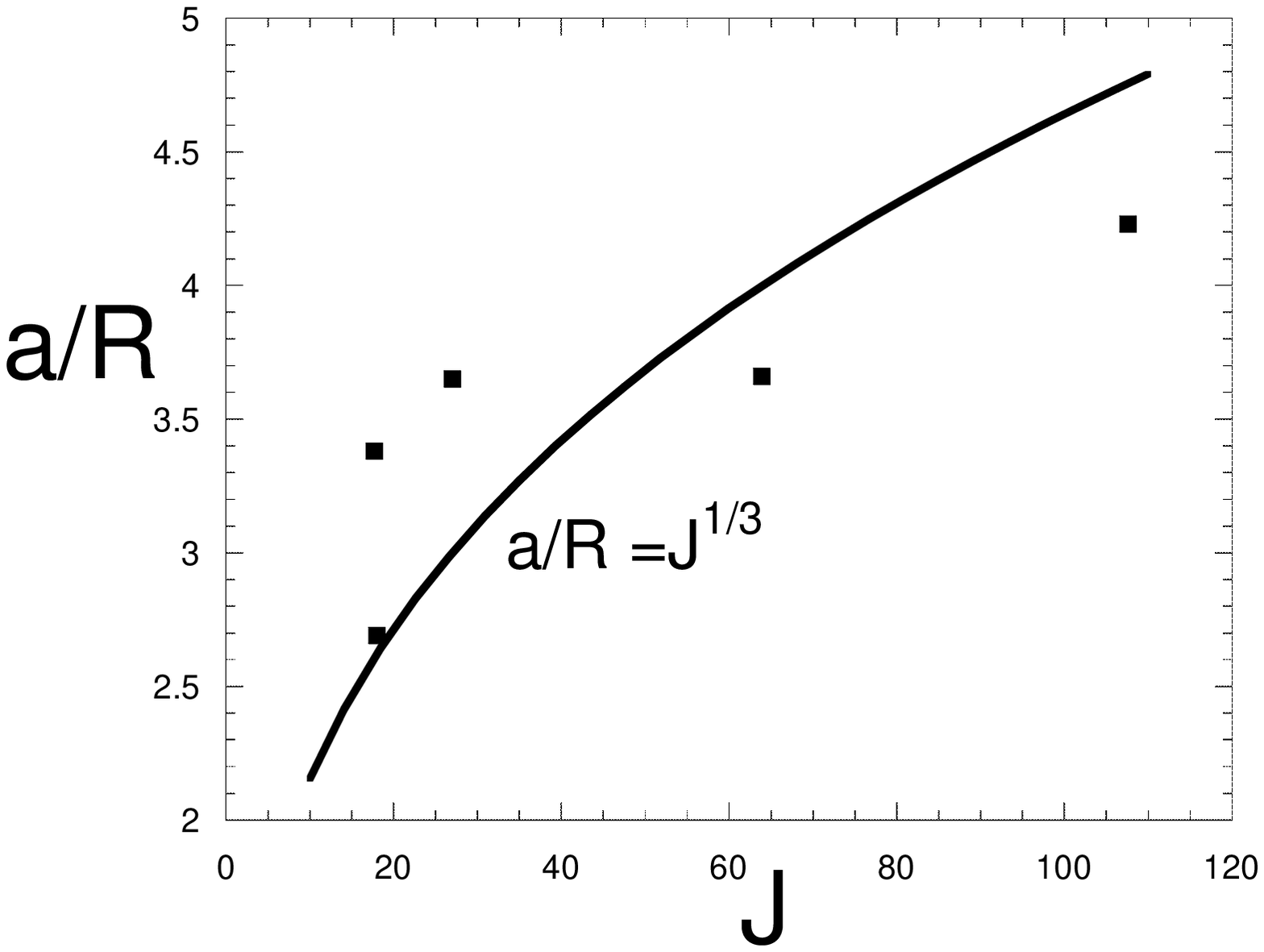} \caption{The
predicted radius of the contact area using $W_D$ from the
Dupr\'{e} formula. Here, $J \equiv 9\pi W /(2 K R)$.  Clearly, the
predicted contact radii are about 2 times smaller.} \label{area1}
\caption{The predicted radius of the contact area using $W$ from
our theory. In contrast to Fig. \ref{area1}, our prediction is
quite good.} \label{area2}
\end{figure}

The equilibrium contact angle $\theta$ follows from minimization of $U_{tot}(h) = U_s(h) + U_{el}(h)$
with respect to $h$.  Figure \ref{contacta} shows $\theta$ as a function of
$2\sqrt{3}\,K R/(\pi \gamma_s)$ for different values of the reduced spreading
parameter $S/\gamma_s$. First, we note that the contact angle is uniquely
determined only if $S$, $\gamma_s$, $R$, and $K$ are specified.  In
the limit $K \rightarrow 0$, we recover the contact angle as obtained from
Young's law but it increases nonlinearly with $K$. This behavior is
consistent with experimental observations.  For $S > 0$,
which corresponds to complete wetting for an ordinary liquid, the contact angle for
the latex particles remains finite for finite $K$.  Moreover, in the limit $S \gg K R $,
the height of the latex particle follows an asymptotic scaling law of
$h \sim R ( K R / S)^{2/3}$.

Our model may provide a way of estimating the adhesion energy $W$. Indeed,
we expect that work is needed to deform the latex particles, and therefore,
some of the gain in the surface energy must be converted into elastic energy.
This implies that the work of adhesion $W$ must be greater than that estimated by the Dupr\'{e}
formula, which only involves the surface tension.  Note that our experiments are
in an intermediate regime where both the surface tension
and elastic energy are important and our model takes both into account.
For an estimate, taking typical values from the latex experiments: $h/R \sim 0.3$,
$R \sim 10^{-7}\,\mbox{m}$, $K \sim 10^6\,\mbox{Pa}$ and
$\gamma_s \sim 50\,\mbox{mJ/m$^2$}$, we find $W_D \sim
100\,\mbox{mJ/m$^2$}$ from Dupr\'{e} formula, and
$W \sim 240\,\mbox{mJ/m$^2$}$ from our model.  Thus, their values
are significantly different with $W$ being twice as big as $W_D$.
To test our model more quantitatively, we
have compared our predictions to the experimental data for the latex
particles as listed in Table \ref{data2}.  The contact angle
is obtained from the experimental values of the height, and it is used to
deduce the spreading parameter $S$ for a given $K$.
The work of adhesion $W$ follows from the relation $W = S + 2 \gamma_s$.
It should be noted that a direct experimental verification of our model
would be to keep $K$, $\gamma_s$, and $W$ fixed, while varying the
particle radius $R$, so that the experimental measurement of the contact angle
could be compared with only one of the curves
shown in Fig. \ref{contacta}.  Unfortunately, in the data
obtained so far, the latex particles come in different sizes; this
direct test will be further explored in the future.  However, using the obtained values of
$W$, we can estimate the contact radii using a formula from the JKR theory:
$a_0/R = \left [ 9 \pi W /(2 K R )\right ]^{1/3}$, and compare this to the experimental
measured contact radius.  The results are shown in Fig. \ref{area1},
where $W$ is estimated from the Dupr\'{e} formula, and Fig. \ref{area2},
where $W$ is obtained from our theory.  It is evident that using our model
to predict $W$ fits the experimental data quite well.

\acknowledgments
We would like to thank L. Mahadevan for fruitful discussions and
Rhodia for financial support and for providing us with the latex
particles.

\end{document}